\documentclass[prd,twocolumn,amsmath,amssymb,nofootinbib,floatfix,superscriptaddress]{revtex4-2}

\usepackage{times,graphics}
\usepackage[colorlinks]{hyperref}
\usepackage{makecell}
\usepackage{diagbox}

\usepackage{graphicx}
\usepackage{dcolumn}
\usepackage{bm}
\usepackage{natbib}
\usepackage{graphicx}
\usepackage{epsfig}
\usepackage{epstopdf}
\usepackage{multirow}
\usepackage{amsmath}
\usepackage{lipsum}
\usepackage{amssymb}
\usepackage{color}
\usepackage{dcolumn}
\usepackage{bm}
\usepackage{natbib}
\usepackage{tabu}
\usepackage{times,graphics}
\usepackage{makecell}
\usepackage{diagbox}
\usepackage{soul,xcolor}
\usepackage{comment}

\RequirePackage{mathptmx}      
\RequirePackage{flushend}

\setstcolor{red}

\graphicspath{{./fig/}}

\begin{document}

\title{A semi-model-independent approach to describe a cosmological database}

\author{Ahmad Mehrabi }
\affiliation{Department of Physics, Bu-Ali Sina University, Hamedan	65178, 016016, Iran}

\begin{abstract}
A model-independent or non-parametric approach for modeling a database has been widely used in cosmology. In these scenarios, the data has been used directly to reconstruct an underlying function. In this work, we introduce a novel semi-model-independent method to do the task. The new approach not only remove some drawbacks of previous methods but also has some remarkable advantages. We combine the well-known Gaussian linear model with a neural network and introduce a procedure for reconstruction of an arbitrary function. In the scenario, the neural network produces some arbitrary base functions which subsequently are fed to the Gaussian linear model. Given a prior distribution on the free parameters, the Gaussian linear model provides a close form for the posterior distribution as well as the Bayesian evidence. In addition, contrary to other method, it is straightforward to compute the uncertainty.
\end{abstract}

\maketitle

\section{Introduction}
A non-parametric approach to describe a database has been originally invented in machine learning scenarios to have as much as possible capacity to describe a complex database. These methods have been widely used in cosmology, specifically in describing the dark energy (DE), since it is an exotic and unknown component in the Universe. In such a scenario, the data has been used directly to reconstruct the underlying function. The Gaussian process (GP) \cite{10.5555/1162254}, the Genetic algorithm (GA)\cite{Bogdanos_2009} and Smoothing method (SM) \cite{Shafieloo_2007} are three most well-known model independent methods which have been used in cosmology. Despite all advantages, these methods have some sever drawbacks. The GP, the most well-known method, is a sequences of Gaussian random variable and has been used for various data in cosmology \cite{Shafieloo_2012,Liao_2019,G_mez_Valent_2018,Pinho_2018,Haridasu:2018gqm,Mehrabi_2020,Bernardo:2021mfs,Briffa:2020qli}. Given a database, the approach gives a mean function and its uncertainty after fitting to the data. A conditional mean and variance are given from a conditional distribution in a GP scenario \cite{10.5555/1162254} and then they have been used to sample a reconstruction. Although the GP provides a straightforward procedure to compute the reconstruction and its uncertainty, it has to sever problems. 1- According to \cite{OColgain:2021pyh}, the GP might provide a smaller parameter's uncertainty compare to some specific cosmological models which is against what we expect from a model independent method. 2- The GP depends on a prior mean function that might affects the final results \cite{Hwang:2022hla}.
        
Moreover, in other two approaches, there is no  generally accepted procedure to estimate the uncertainty. In fact, this is the most important drawback of these methods. The GA uses some base functions and try different combination of those to create a suitable reconstruction. The method has been used to describe different cosmological data in \cite{Nesseris_2010,Nesseris_2012,Arjona:2020axn,Arjona:2020skf}. In this scenario, there is no clear and straightforward way to estimate the uncertainty. To address this, the $\chi^2$ distribution has been used in \cite{Vazirnia:2021xuu} and path integral in \cite{Nesseris_2012,Nesseris:2013bia} to estimate the uncertainty.
      
In the SM scenario, the method starts from an initial guess and a iterative process has been used to find a curve with lower $\chi^2$. Similar to the GA, the approach does not provide a reliable procedure to estimate the uncertainty. The SM has been used in various investigation of cosmological data in \cite{Shafieloo_2007,Shafieloo_2010,Shafieloo_2018,Vazirnia:2021xuu,Mehrabi_2022}.      

On the other hand, in a model dependent scenario, a parametric model with some free parameters has been used to describe a database. Given a data base and model, one can easily build a likelihood function, then using a statistical viewpoint like Bayesian, it is straightforward to do the parameter inference.  The Gaussian linear model (GLM) is a simple model which is linear in its parameters and can be used to describe a database \cite{Trotta:2017wnx}. Given a Gaussian prior on the free parameters, the method provides an analytic formula for posterior distribution as well as the Bayesian evidence. In fact, this is the most advantage of the GLM because numerical estimation of these quantities are computationally expensive. Notice that the evidence is a crucial quantity in Bayesian model selection and can be used to falsify a model using a database.
Notice that the GLM needs some base functions which should not be depends on the free parameters. Since these functions can be quite non-linear,  it is possible to create them through a neural network (NN). In current work, we combine the GLM approach with an NN and introduce a novel approach to describe a database. The method, not only gives a close form for posterior distribution and evidence but also provides a reasonable way to estimate the uncertainty.

The structure of this paper is as follows:  In Sec.~\ref{sec:glm}, we give the basic formalism of the GLM and provide the analytic formula to obtain the posterior distribution as well as the evidence. In Sec.~\ref{sec:net}, the details of a simple NN is given and then the procedure of our methods is described. Then in Sec.~\ref{sec:obs}, we consider recent Hubble as well as SNIa data and apply the method to find the best reconstruction.  Moreover, the databases have been used to constrain the benchmark $\Lambda$CDM model and the results are compered with the reconstruction. Finally, we conclude and discuss the main points of our finding in section Sec.~\ref{conclude}.

\section{Gaussian linear model}\label{sec:glm}
In this section, we briefly review the basic formalism of the GLM. In this scenario, a database is modeled by a linear combination of some base functions
\begin{equation}\label{eq:model}
f(x,\theta) = \sum \theta_j X^j(x),
\end{equation}   
where $X^j(x)$ are the base functions and each can be an arbitrary non-linear function of x. The $\theta_j$s indicate the free parameters. Given a database as $(x_i,y_i,\tau_i)$, the likelihood function is given by
 \begin{equation}\label{eq:likelihood}
 p(y|\theta) = \mathcal{L}_0\exp[-\frac{1}{2}(\theta-\theta_0)^tL(\theta-\theta_0)]
 \end{equation}
 where,
\begin{equation}\label{eq:likelihood-nor}
    \mathcal{L}_0= \frac{1}{(2\pi)^{n_{obs}/2}\Pi\tau_i} \exp[-\frac{1}{2}(b-A\theta_0)^tL(b-A\theta_0)],
\end{equation}
 and 
\begin{equation}\label{eq:L-theta_0}
    F_{ij}=X^j(x_i)~~,~~A= \frac{F_{ij}}{\tau_i}~~,~~b=\frac{y_i}{\tau_i}~~,~~L=A^tA~~,~~\theta_0=L^{-1}A^tb.
\end{equation}
Here the maximum likelihood occurs at $\theta_0$ and $L$ indicates the Fisher matrix. 
 
In order to perform the Bayesian parameter inference, we need to define a prior on the free parameters. We consider a Gaussian prior as
\begin{equation}\label{eq:prior}
p(\theta) = \frac{|\Sigma_{pri}|^{-1/2}}{(2\pi)^{n_{par}/2}}\exp[-\frac{1}{2}(\theta-\theta_{pri})^t\Sigma_{pri}^{-1}(\theta-\theta_{pri})],
 \end{equation}
 where $\theta_{pri}$ ($\Sigma_{pri}$) is the mean (covariance matrix) of the prior and $n_{par}$ denotes the number of free parameters. Using the Bays theorem, the posterior distribution is proportional to
  \begin{equation}\label{eq:pos}
  p(\theta|y) \propto \exp[-\frac{1}{2}(\theta-\theta_{pos})^t\Sigma_{pos}^{-1}(\theta-\theta_{pos})],
  \end{equation}
  where $$\Sigma_{pos}^{-1} = \Sigma_{pri}^{-1} + L $$ and $$\theta_{pos}=(\Sigma_{pri}^{-1} + L)^{-1}(\Sigma_{pri}^{-1}\theta_{pri}+L\theta_0)$$
  From above posterior distribution, it is straightforward to compute the mean value of parameters as well as their uncertainties.
  
In addition, the GLM provides the Bayesian evidence which is a key quantity in model selection. The Bayesian evidence is given by
  \begin{equation}\label{eq:evid_gen}
p(y) = \int d\theta p(\theta)p(\theta|y).
\end{equation}
Since both the prior and likelihood are a multivariate Gaussian, the integral has an analytical solution.
The close form of the Bayesian evidence is given by, 
  \begin{equation}\label{eq:evid}
  p(y) = \mathcal{L}_0|\Sigma_{pri}|^{-1/2}|\Sigma_{pos}|^{1/2}\exp(D),
  \end{equation}
  where 
  \begin{eqnarray}\label{eq:D}
  D &=& \frac{1}{2}[(\theta_{pri}^t\Sigma_{pri}^{-1} + \theta_{0}^tL)(\Sigma_{pri}^{-1}+L)^{-1}(\Sigma_{pri}^{-1}\theta_{pri}
   +L\theta_0) \nonumber\\ &-& (\theta_{pri}^t\Sigma_{pri}^{-1}\theta_{pri} + \theta_{0}^tL\theta_{0})],
  \end{eqnarray} 
 and $|\Sigma|$ denotes the determinant of $\Sigma$.
 
 Since in the GLM the base functions are quite arbitrary, in current work, we are going to generate them through an NN and then find the best reconstruction.  

\section{Neural network}\label{sec:net}
\begin{figure}[h]
	\centering
	\includegraphics[width=0.5\textwidth]{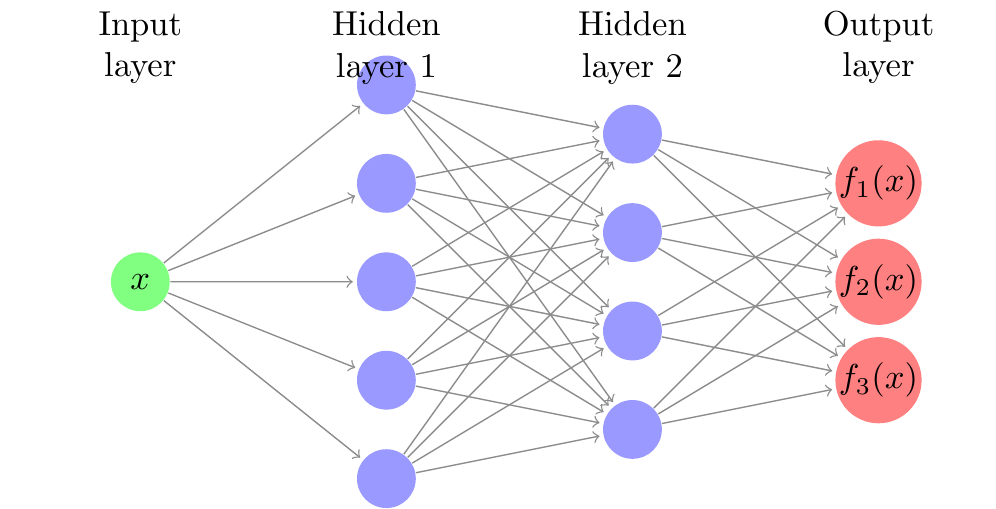}	
	\caption{A simple NN with two hidden layers which gives three arbitrary functions.  }
	\label{fig:net}
\end{figure}

A simple NN is consisted of some hidden layers each with some arbitrary number of neurons. Each layer applies a linear transformation as $y^l=\hat{y}^{l-1}A^T+b$ to the incoming data $\hat{y}^{l-1}$. Then an activation function applies to the output $\hat{y}^l=Act(y^l)$ and send to the next layer. In this formalism, "A" is a matrix containing the weights and "b" indicates bias term.  In Fig.(\ref{fig:net}), a simple NN with two hidden layers has been shown. The first (second) layer has 5 (4) neurons. The input is x and the NN gives 3 arbitrary function of x. By changing the number of hidden layers and neurons, it is possible to generate very complex function. We use this property to generate the base functions of the GLM scenario and hence the best reconstruction. In fact, our new approach is mainly consisted of two parts, 1- Finding the base functions through an NN which is quite model-independent and 2- The GLM framework, which has an analytic solution. This is the reason why we name the method a semi-model-independent approach.

Given a database, one can fix the number of hidden layers and neurons, then using the $\chi^2$ as loss function, it is possible to find a set of suitable base functions (which give a minimum $\chi^2$). Finally, the database is modeled by these base functions through a GLM scenario.
For the activation function, we examine different options including, Gaussian Error Linear Unit (GELU), Exponential Linear Unit (ELU), Leaky Rectified Linear Unit (leaky RELU) and Sigmoid Linear Unit (SiLU). It is worth noting that the final results are independent of the activation choice and it only affects the speed of convergence. Throughout this work, we use SiLU activation because it gives a faster convergence compare to other options. To make this procedure as transparent as possible, we made a github Repo and publish all code. (see \href{https://github.com/Ahmadmehrabi/GLM_Network}{This Repo}).   

\section{Applying to the cosmological data}\label{sec:obs}
 In this section, we apply the method on two well-known cosmological data, namely the Hubble and SNIa databases. The Hubble database is consisted of the cosmic chronometer and radial BAO collected in \cite{Farooq:2016zwm}. We also add the current $H_0$ measurement (the SHOES data)\cite{Riess:2019cxk}.  These data points give a direct measurement of the Hubble parameter up to redshift $z\sim 2.5$. On the other hand, we consider the recent SNIa data from Pantheon sample \cite{Scolnic_2018}, which contains 1048 spectroscopically confirmed SNIa up to redshift $z=2.26$. 
 
  For each database, we perform the following steps,
\begin{itemize}
	\item Construct an NN with certain numbers of hidden layers and neurons
	\item Fix number of base functions
	\item Train the network using $\chi^2$ as a loss function and find the minimum $\chi^2$ (after some epochs the $\chi^2$ does not decrease significantly)
	\item Freeze the wights and biases of each hidden layers to generate a set of suitable base functions
	\item Consider a wide Gaussian prior on free parameters (notice that the number of parameters equals the number of base functions)
	\item Apply the GLM formalism to find the best value of the parameters and their uncertainties
	\item Use these parameters and their uncertainties to find the best reconstruction as well as its uncertainty
\end{itemize}
 
\begin{figure}[h]
	\centering
	\includegraphics[width=0.5\textwidth]{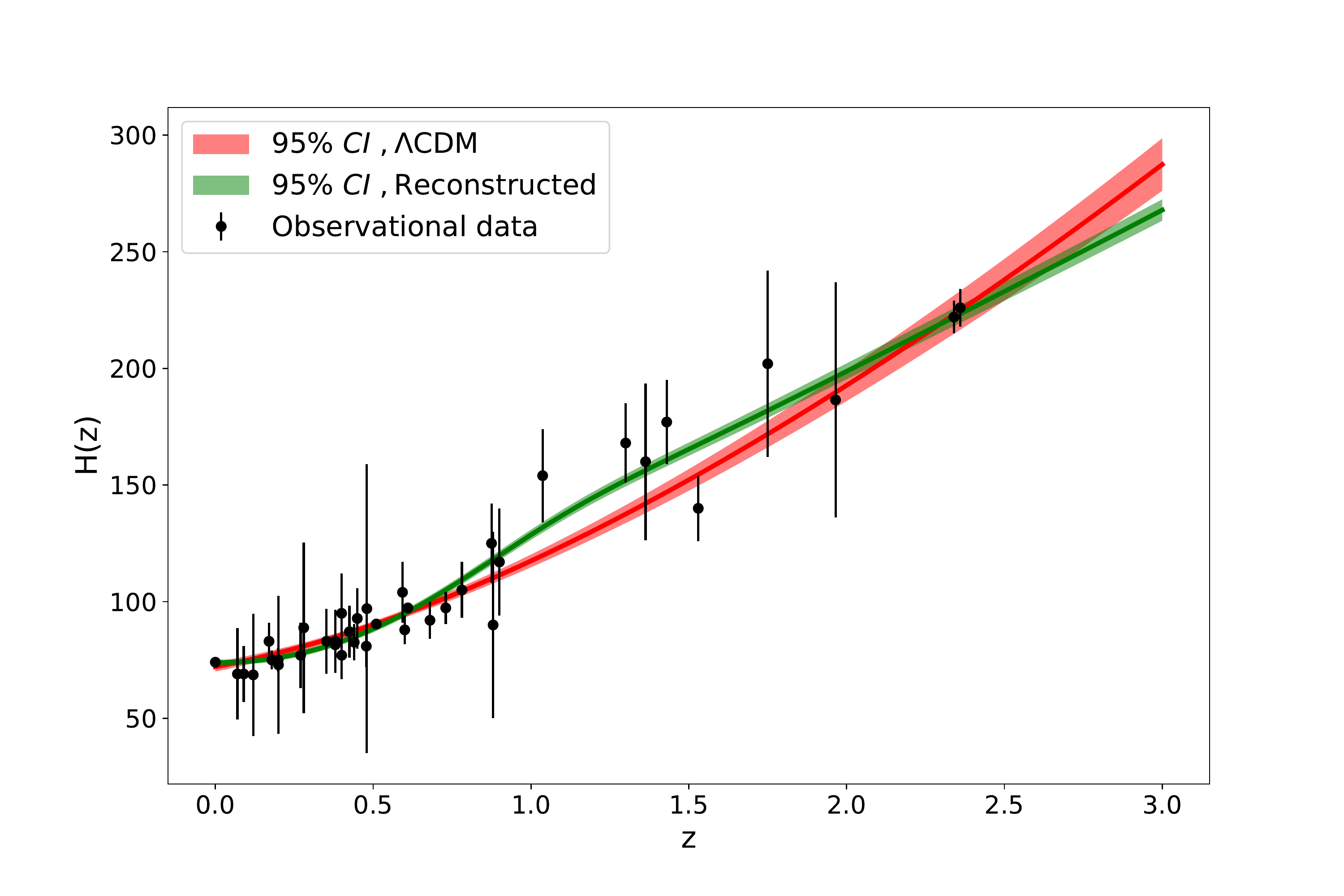}	
	\caption{The green color: reconstructed and its $95\%$ confidence interval (CI) considering only one base function. The red color: the best fit and $95\%$ CI in the $\Lambda$CDM model. The observational data are the Hubble data explained in the text. }
	\label{fig:hub1}
\end{figure}

  For the Hubble data, we consider two hidden layers with 5, 3 neurons respectively and examine two scenarios . In the first, only one base function has been considered so we have only one free parameter. The results is depicted in Fig(\ref{fig:hub1}).

For the low redshift data $z<1$, both the reconstruction and $\Lambda$CDM provide almost the same results. On the other hand, there are some data points in redshift range  $z\in (1-2)$ which clearly the $\Lambda$CDM is not able to describe very well. Contrary, the best reconstruction provides a better fit for the data points at this redshift interval. Notice that both approaches give a similar results for data points at $z\sim 2.5$. It is clear that our method shows a more flexibility in describing a database and can provide a reconstruction with smaller value of $\chi^2$. In fact, the best reconstructed function gives $\chi^2_{min}\sim 17 $ which is considerably smaller than the best $\Lambda$CDM ($\chi^2_{min}\sim 31$) for the database.
  
\begin{figure}[h]
	\centering
	\includegraphics[width=0.5\textwidth]{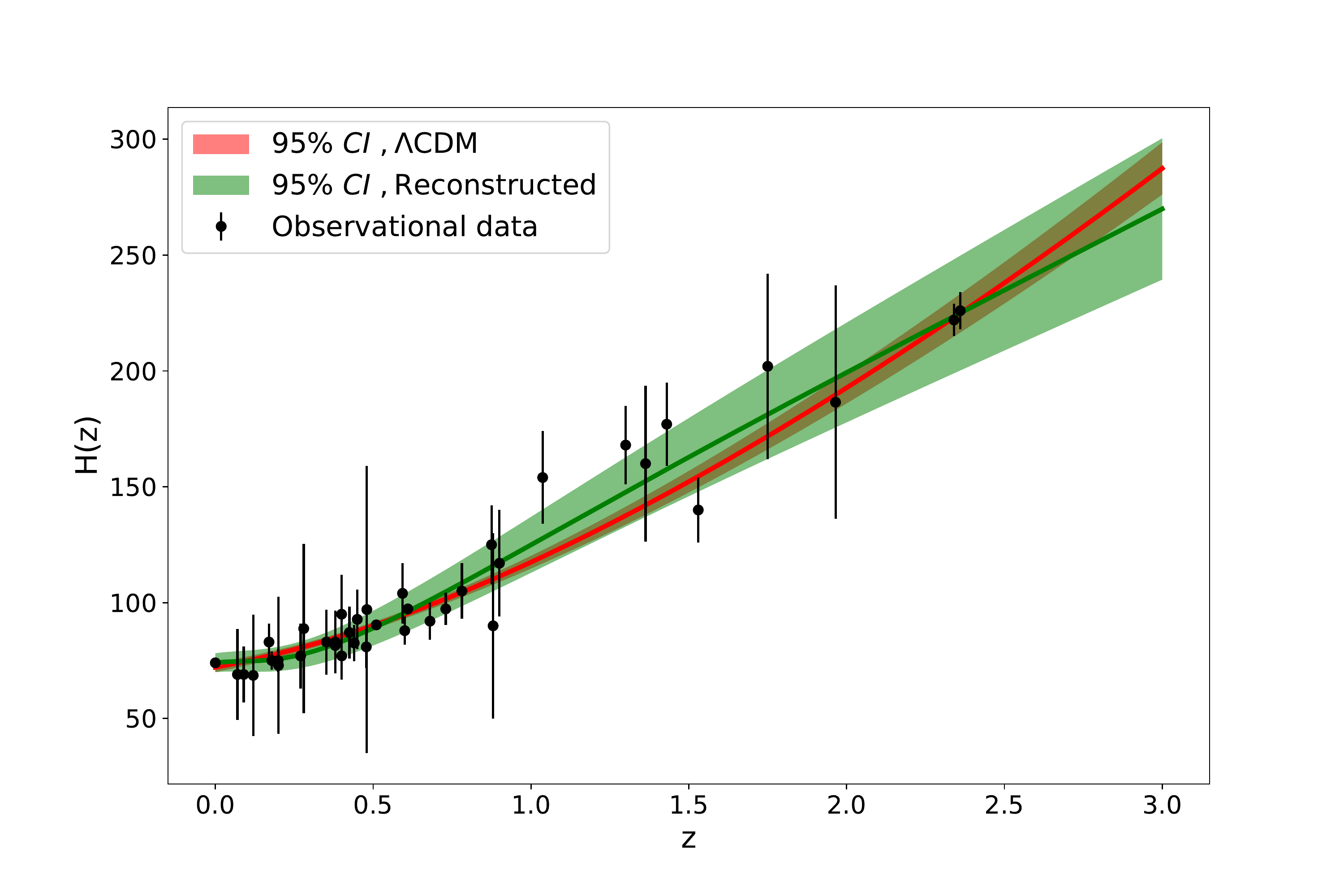}	
	\caption{The green color: reconstructed and its $95\%$ CI considering two base functions. The red color: the best fit and $95\%$ CI in the $\Lambda$CDM model}
	\label{fig:hub2}
\end{figure}

In the second scenario, we consider two base functions with a similar NN architecture. The result is shown in Fig.(\ref{fig:hub2}). The best reconstructed function is almost the same as previous case but more free parameters provides a larger uncertainty. Notice that in a model dependent scenario, a model with more free parameters gives a larger uncertainty. The results indicate that one should examine different number of base functions to find a reasonable uncertainty for the reconstruction.  

\begin{figure}[h]
	\centering
	\includegraphics[width=0.5\textwidth]{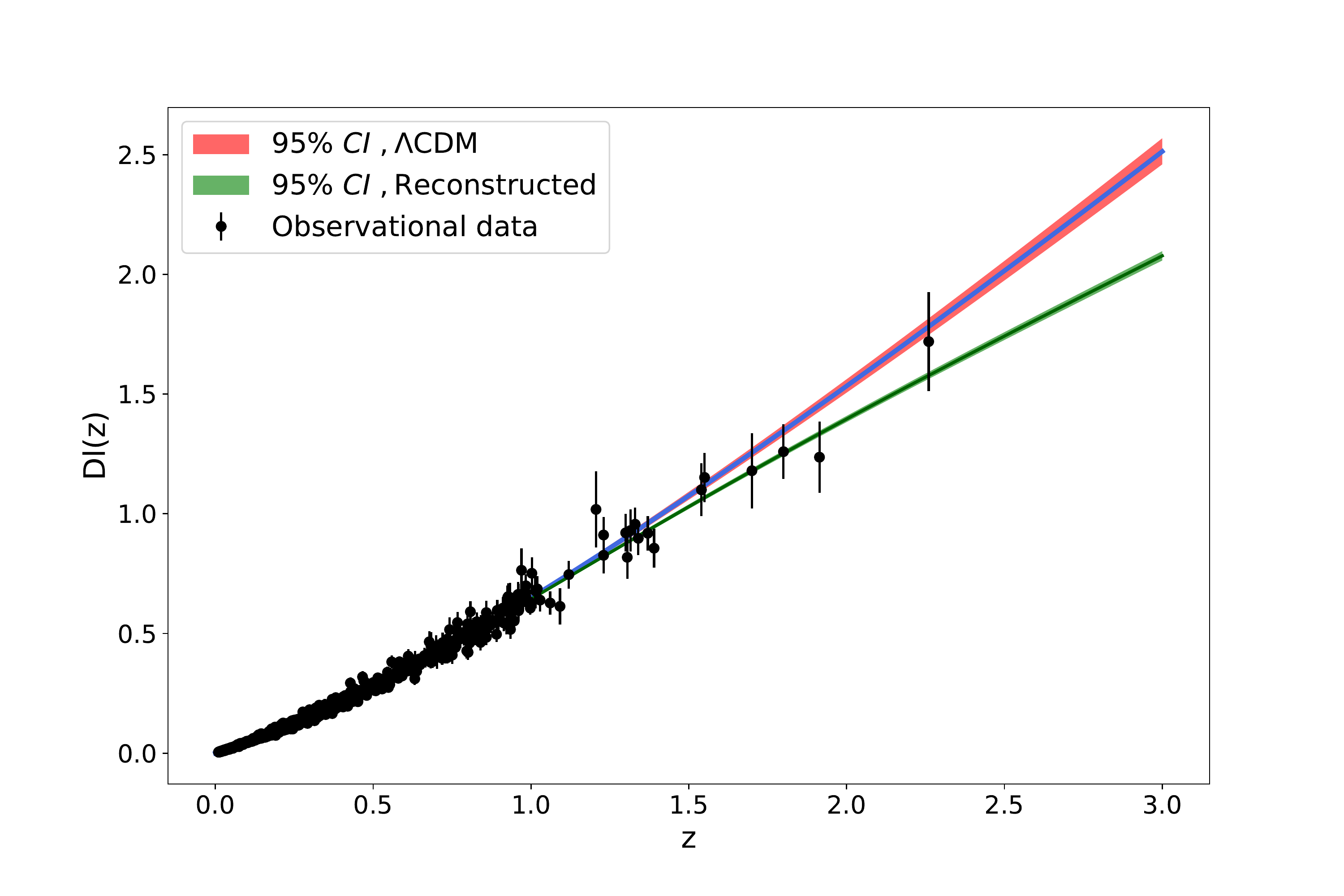}	
	\caption{The green color: reconstructed and its $95\%$ CI considering two base functions. The red color:  the best $\Lambda$CDM model and its $95\%$ CI. The observational data are the luminosity distance from Pantheon sample. }
	\label{fig:Dl2}
\end{figure}

\begin{figure}[h]
	\centering
	\includegraphics[width=0.5\textwidth]{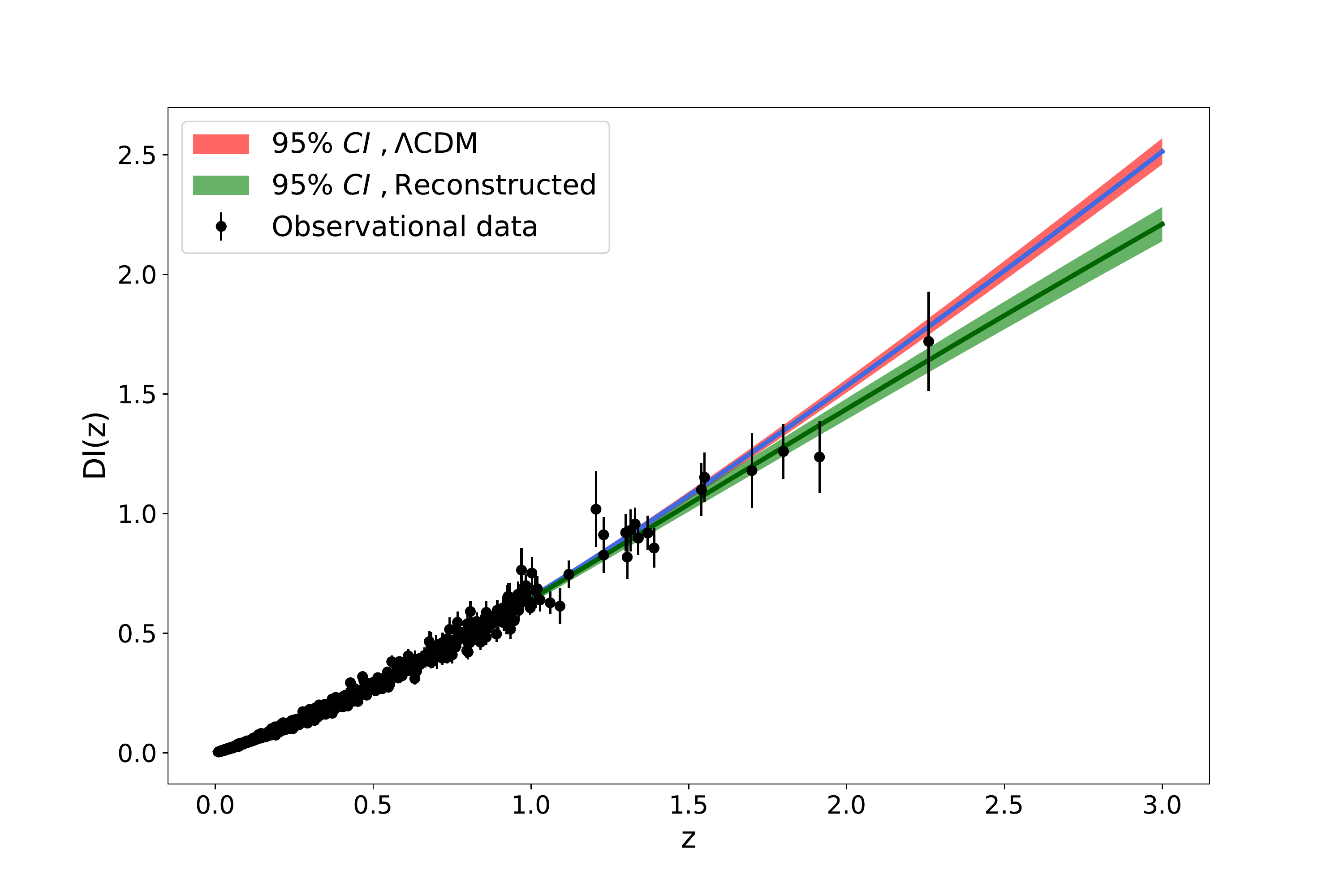}	
	\caption{The green color: reconstructed and its $95\%$ CI considering three base functions. The red color: the best $\Lambda$CDM and its $95\%$ CI. The observational data are the luminosity distance from Pantheon sample.}
	\label{fig:Dl3}
\end{figure}

In order to check the reliability of our method, on the other hand, we apply it to the luminosity distance of the SNIa database. Considering the Pantheon sample \cite{Scolnic_2018}, the absolute magnitude set to $M=-19.3$ to convert the module distance to the luminosity distance. Similar to the previous case, we consider a two hidden layers network but this time with 10 and 6 neurons respectively. In this case, it is difficult to find a reconstruction with only one base function so we perform the procedure using 2 and 3 base functions. The results for these two cases have been shown in Fig.(\ref{fig:Dl2}) and (\ref{fig:Dl3}) respectively. In the first case, we see a very small uncertainty which is even smaller than the $\Lambda$CDM. The best reconstruction gives $\chi^2_{min}\sim 1009$ which is relatively smaller than the best $\Lambda$CDM ($\sim 1031$). The $\Lambda$CDM and the reconstructed function is almost the same up to redshift $z\sim 1.5$ but there are some high redshift data points which the reconstruction provides a better fit so gives a smaller $\chi^2$. Notice that, the $\Lambda$CDM is not flexible enough to describe these high redshift data points but our method can easily provides a fit which is much better than the $\Lambda$CDM. 

Similar to the Hubble data, number of base functions determines the uncertainty of the reconstruction. The results in Fig.(\ref{fig:Dl3}) indicates that the 3 base functions provides more reasonable uncertainty for the luminosity distance. On the other hand, changing the number of base function does not change the best reconstruction significantly. It is worth noting that, these results are consistent with those obtained in \cite{Vazirnia:2021xuu} which used the same data but different model independent methods. In fact, as pointed out in \cite{Vazirnia:2021xuu}, the best reconstruction using GP (GA) gives 1010.2 (1008.4) for the Pantheon data set, hence our new method is able to find the best reconstruction as well as other well-known approaches. Moreover, our method is much less computationally expansive compared to other methods specially SM and GA. 

Finally, we summarize main points of our method and results in this section:
\begin{itemize}
	\item The combination of an NN and GLM provides a framework to investigate a database in a model-independent manner
	\item The NN provides a set of suitable base functions for the GLM part
	\item The GLM gives a close form for the posterior distribution of the free parameters
	\item The method provides a reliable way to compute the uncertainty which depends on the number of base functions
	\item To find a reliable uncertainty, one should examine different number of base functions 
	\item The best reconstruction is consistent with other model-independent methods considering both the Hubble and SNIa data

	\item The method is much less computationally expansive compered to other well-known methods
\end{itemize}

\section{Conclusion}\label{conclude}
In this work, we introduce a new semi-model-independent method to reconstruct a function directly from a database. The method is the combination of an NN and GLM approach. The NN part generates some arbitrary base functions which are fed to the GLM to find a suitable reconstruction. Since our method is a combination of a model-independent (NN part) and a part with analytic solution (the GLM), we name it the semi-model-independent. In this scenario, the NN tries to find a set of suitable base functions which gives a minimum value for the $\chi^2$ and then the wights and biases are frozen to have some fix base functions. Finally, the GLM uses the base functions and gives the best value of parameters (which are coefficient of the base functions) as well as their uncertainties. In the last stage, we consider the Bayesian statistic to infer the free parameters. Given the parameter's uncertainty, it is straightforward to obtain the uncertainty of the reconstructed function. 

We apply the method on both recent Hubble and SNIa database. For both cases, we adopt an NN with two hidden layers. Considering the Hubble data, the method finds a reconstruction with a $\chi^2$ which is $\sim 14$ unite smaller than the best $\Lambda$CDM. Moreover, examining different number of base functions, we find out that  having two base functions is reliable for the Hubble data. Our results indicate that the best reconstruction is much better in describing data points specifically at redshift $z\in (1-2)$ range compered to the $\Lambda$CDM. 

In addition, we apply the method on the SNIa data to investigate its performance. In this case, first the module distance is converted to the luminosity distance and then fed to the method in order to find the best reconstruction. We perform the reconstruction with 2 and 3 base functions. While the best reconstruction is almost the same, the 3 base functions gives more reliable uncertainty. We observer that the method gives a better fit compared to the $\Lambda$CDM specifically for data points in redshift $z\in (1.5-2)$ range. The method provides a function with $\chi^2\sim 1009$ which is 21 unite smaller than the best $\Lambda$CDM. Moreover, compere to the other model-independent approaches, our method is much less computationally expansive while the best reconstruction does not change significantly. 
    
To make this procedure as transparent as possible, we made the code for current work available for others to use and improve upon\footnote{\url{https://github.com/Ahmadmehrabi/GLM_Network}}.



\bibliographystyle{apsrev4-1}
\bibliography{ref}

\end{document}